\documentstyle[preprint,eqsecnum,aps,floats]{revtex}
\begin{document}

%
\begin{titlepage}
\title{Born--Infeld action for gravitational and electroweak  fields}
\author{Dmitriy Palatnik\thanks{mailto: palatnik@ripco.com}
}

\maketitle

\tighten
\begin{abstract}

This note suggests a generalization of the Born--Infeld action (1932)
on the case of electroweak and gravitational fields in 4-dimensional spacetime.
Basic notions one deals with
are Dirac matrices, $\gamma_{a}$, and dimensionless covariant
derivatives, $\pi_{a} = - i\ell \nabla_{a}$.
The action, constructed from $\gamma_a$ and $\pi_a$, contains a parameter,
$\ell$, (which is of order of magnitude of Planck's length), 
and possesses additional symmetry with respect to transformations
of the Lorentz group imposed on pairs ($\gamma_{a}$, $\pi_{a}$).
This symmetry leads to the most natural coexistence between standard electroweak
and gravitational action terms in the action's expansion in $\ell$, and provides
basis for unification of the interactions.
It's shown that the parameter of the Lorentz group is associated with
a constant value of the electroweak potential at spatial infinity.

\end{abstract}

\vspace{20pt}
\center{PACS numbers: 11.10.Lm; 12.10.-g; 12.60.-i; 11.25.Mj}

\end{titlepage}

\tighten

\section{Introduction}

\vspace{10pt}
\hfill\parbox{9cm}{%
\small The present educated view of the standard model, and of
general relativity,
\small is ... that these are the leading terms in effective field theories.\\
\mbox{}\hfill\small S.~Weinberg \cite{weinberg}
}
\vspace{14pt}

Development in string models has led  to rediscovery
of  the Born--Infeld (BI) theory \cite{BI,GWG,k}, initially formulated for
electromagnetic fields. It was shown \cite{ft,bs,mr,ts}, that
in the theories of open strings the low-energy effective actions
coincide with the BI one. As a result, research in both electromagnetic 
and gravitational BI gained new momenta.

Nonlinear modifications of the standard gravitational and electromagnetic
actions have been studied with different corrections. For example,  couplings
of the BI electromagnetic action to the Einstein--Hilbert (EH) one 
have been considered by Gibbons and Herdeiro \cite{gh},  by
Wirschins et al. \cite{wsj},
by Ay\'on-Beato and Garc\'{\i}a \cite{abg1,abg2,abg3},  by Chang
et al. \cite{cmm}.
Nonlinear in curvature tensor gravitational actions of BI type
have been studied by
Feigenbaum et al. \cite{ffp}, by Wohlfarth \cite{w}, by Nieto \cite{n}, and also in \cite{9701017}.

The motivation for original BI theory was the possibility to obtain finite energy solutions to
the BI field equations, which should describe, according to the authors,  structure of an electron. 
Having that in mind, one attempts to construct from scratch (i.e., with no reference to any string
model) a BI action for gravitational and electroweak fields (`BI-gew'), which should, hopefully, lead to
solutions, corresponding to field condensations with finite energy.

An essential feature of BI-gew, suggested below, 
is nonlinearity of the  action in curvature tensor, $\rho_{ab}$,
which substitutes, roughly speaking, dimensionless electromagnetic tensor, $F_{ab}$, 
in BI invariant, $\int d\Omega \sqrt{-\det(\eta_{ab} + F_{ab})}$.
Due to this nonlinearity one obtains both linear and quadratic in $\rho_{ab}$
terms in the action's expansion, corresponding, respectively, to the EH 
term for gravity and to the standard one for electroweak fields. 

As an asset of BI-gew (as well as of original BI), one may consider the fact
that action invariants contain only one dimensional parameter, $\ell$, 
which fixes coefficients of the action's expansion in powers of the curvature tensor.

One may specify yet another reason for considering the BI-gew action;
namely, its additional (to general covariance, gauge invariance, and invariance under
matrix transformations) symmetry with respect to transformations of Lorents group
imposed on pairs  ($\gamma_a$, $\pi_a$),\footnote{{ }%
This symmetry is analogous to the symmetry of the (1-dim) action for a  relativistic particle,
$-\int \sqrt{dt^2 - dx^2}$, invariant with respect to Lorents transformations imposed
on pair ($dt$, $dx$). One may observe, that the relativistic particle action was the 
inspiration for original BI one.}
where $\gamma_a$ are the Dirac matrices and $\pi_a$ are dimensionless covariant derivative operators.
The symmetry is introduced `by hands' and makes BI-gew look more appealing aesthetically.
Besides aesthetical appeal, one may offer additional argument in favor of  selecting a more symmetric action. 
The higher is the symmetry of the action, the more restricted is the class of invariants.
Since the goal of modern research is to find the unified field theory with unique action invariant(s), one 
should welcome any upgrade of existing theory, leading to higher symmetry of the action, 
than that of its predecessor.

One is able  to conceal all interactions in connections,
associated with $\pi_a$. As a result all interactions
enter the action on equal footing. One believes, that if
unification of the interactions is feasible in four dimensions at all,
then given approach may provide basis for such unification.

To begin with, one specifies basic elements of the action,
compatible with quantum field theory in 4-dimensional spacetime;
namely,   ($\alpha$) Dirac
$N \times N$ matrices, $\gamma_{a}$, submitted to relations
\begin{eqnarray}\label{gdef}
\gamma_{(a}\,\gamma_{b)} & = & g_{ab}\hat{1}_N\,,
\end{eqnarray}
where $g_{ab}$ is the metric tensor, and $\hat{1}_N$ is unit $N \times N$ matrix; ($\beta$)
dimensionless operators
\begin{equation}\label{pi}
\pi_a \,=\, -i\ell\nabla_a\,,
\end{equation}
where $\nabla_a$ is a covariant derivative and $\ell$ is a
parameter. Action of (\ref{pi}) on spinors, scalars, and Dirac matrices
is specified in the following text. For spinors and scalars,
for example, one obtains,
\begin{eqnarray}\label{pipsi}
\pi_a\Psi  & = &  -i\ell\left({\partial}_a\Psi - \Gamma_{a}\Psi\right)\,;\\
\label{piphi}
\pi_a\Phi  & = &  -i\ell\left({\partial}_a\Phi - \dot\Gamma_{a}\Phi\right)\,;
\end{eqnarray}
here $\Psi$ ($\Phi$) represents spinor (scalar), and $\Gamma_a$  ($\dot\Gamma_a$)
are connection matrices. Introducing $\Psi_a = \gamma_a\Psi$, one obtains,
\begin{eqnarray}\label{pipsia}
\pi_b\Psi_a  & = &  -i\ell\left({\partial}_b\Psi_a - \Gamma_{ab}^c\Psi_c - \Gamma_{b}\Psi_a\right)\,.
\end{eqnarray}
Here $\Gamma_{ab}^c = \Gamma_{(ab)}^c$ is connection symbol.\footnote{{ }%
The connection symbol, $\Gamma_{ab}^c$, coincides with the Christoffel one only in leading
approximation in $\ell$.  One's intent is to use the Palatini method in order to find $\Gamma$'s 
and $\gamma$'s.}
From (\ref{pipsi}) and (\ref{pipsia}) one may easily obtain $[\pi_b, \gamma_a]$.
By a postulate, the action is form-invariant with respect to substitutions,
$\gamma_a \mapsto \gamma_a'$,  $\pi_a \mapsto \pi_a'$, where
\begin{eqnarray}\label{symm}
      \gamma_a' & = &
\cosh\theta\,\gamma_{a} + \sinh\theta\,\pi_a\,,\\
\label{symm1} \pi_a' & = &
\sinh\theta\,\gamma_{a} + \cosh\theta\,\pi_a\,,
\end{eqnarray}
and $\theta$ doesn't depend on coordinates.
In section VI the physical meaning of parameter $\theta$ is clarified.
It is shown that $\theta$ is associated with the value of electroweak
potential at spatial infinity.

\section{Objects of the theory}

1. One may introduce dimensionless curvature tensor,
\begin{eqnarray}\label{curvature}
{\rho}_{ab} & = & 2\pi_{[a}\pi_{b]}
       =  2\ell^2\left(\partial_{[a}\Gamma_{b]} -
\Gamma_{[a}\Gamma_{b]}\right)\,,
\end{eqnarray}
where $\Gamma_a$ are connections, given in spinorial or
scalar representation.
Operator $\rho_{ab}$ is matrix $N \times N$.

2. Define a tensorial operator,
\begin{eqnarray}\label{phiop}
\phi_{ab} & = & \gamma_{[a}\gamma_{b]} - \pi_{[a}\pi_{b]}\,.
\end{eqnarray}

3. Define a scalar density,
\begin{eqnarray}\label{phi}
\phi & = & {1\over
{5!\,N}}\,e^{abcd}\,e^{efgh}\,Tr\left\{\phi_{ae}\,
\phi_{bf}\,\phi_{cg}\,\phi_{dh}\right\}\,,
\end{eqnarray}
where $e^{abcd} = e^{[abcd]}$ is the absolute antisymmetric symbol,
$e^{0123} = 1$.

4. Define tensorial differential operators, 
\begin{eqnarray}\label{chi1}
\chi_{ab} & = & \gamma_{(a}\gamma_{b)} - \pi_{(a}\pi_{b)}
\, = \,  g_{ab} - \pi_{(a}\pi_{b)}\,.
\end{eqnarray}
Here one used (\ref{gdef}).

5. Define a scalar density (differential operator),
\begin{eqnarray}\label{detchi}
\chi & = & {1\over
{4!}}\,e^{abcd}\,e^{efgh}\,\left\{\chi_{ae}\,
\chi_{bf}\,\chi_{cg}\,\chi_{dh}\right\}\,.
\end{eqnarray}

6. Define tensorial differential operators, $\chi^{ab}$, satisfying relations,
\begin{eqnarray}\label{chi1inv}
\chi^{ab}\,\chi_{bc} & = & \delta^a_c\,.
\end{eqnarray}
One may easily find, $\chi^{ab} = g^{ab} + \pi^{(a}\pi^{b)} + O(\ell^4)$, 
where $\pi^a = g^{ab}\pi_b$.

Scalar densities, reducing to $\sqrt{-g}$ in the limit
$\ell\rightarrow0$, are $\sqrt{-\phi}$ and $\sqrt{-\chi}$.\footnote{{ }%
One should understand  $\sqrt{-\chi}$ as 
 respective expansion in powers of $\ell^2$.}
Objects, defined in (\ref{phiop}) -- (\ref{chi1inv}),
are form-invariant with respect to $\theta$-rotations (\ref{symm}) and (\ref{symm1}).

One may prove a useful formula,
\begin{eqnarray}\label{myform}
e^{abcd}\,e^{efgh}\,\gamma_{[a}\gamma_{e]}\,\gamma_{[b}\gamma_{f]}\,
\gamma_{[c}\gamma_{g]} & = & 10\,g\,\gamma^{[h}\gamma^{d]}\,,
\end{eqnarray}
following from (\ref{gdef}).\footnote{{ }Here $\gamma^a =
g^{ab}\,\gamma_b$, where $g^{ab}g_{cb} = \delta^a_c$.}

\section{Action for the fields}

1. Consider a system including electron, neutrino, electroweak
and gravitational fields, together with scalar fields (quadruplets) $U$ and $V$.
Introduce two spinorial bases. First, $L$-basis, is represented by
octets ($N  = 8$),
$L_{el}   =  \lbrack \nu_L\,, e_L\rbrack^T$,
$\overline{L}_{el} = \lbrack \overline{\nu}_L\,,
\overline{e}_L\rbrack$.
Here $e_L$ and $\nu_L$ are 4-spinors for the left-handed electron and
neutrino, respectively; subscript $el$ reminds that
respective spinors belong to the electron's flavor.
The second, $R$-basis, is represented by quadruplets, ($N  = 4$),
$e_R$ and $\overline{e}_R$,
corresponding to a right-handed electron.

2. One may specify the action of $\pi_a$ on introduced fields.
\begin{eqnarray}\label{cderiv1}
\pi_aL_{el} & = & -i\ell\left(\partial_a +
{i\over2}g^{\prime}\,\sigma_k\otimes\hat1_4\,W_a^k
+ {i\over 2}g^{\prime\prime}\,{Y}_L\otimes\hat1_4\,B_a -
\hat1_2\otimes{\overline\Gamma}_a\right)L_{el}\,,\\
\label{cderiv2}\pi_ae_R & = & -i\ell\left(\partial_a +
{i\over 2}g^{\prime\prime}\,{Y}_R\,B_a -
{\overline\Gamma}_a\right)e_R\,,\\
\label{cderiv3}\pi_aU & = & -i\ell\left(\partial_a +
{i\over2}g^{\prime}\,\hat{1}_2\otimes\sigma_k\,W_a^k
+ {i\over 2}g^{\prime\prime}\,{Y}_W\,B_a
\right)U\,,\\
\label{cderiv4}\pi_aV& = & -i\ell\,\partial_a V\,,
\end{eqnarray}
where $B_a$ and $W_a^k$ ($k = 1,\,2,\,3$) are potentials of
electroweak fields; ${Y}_L = -\hat{1}_2$,
${Y}_R = -2\times\hat1_4$, and $Y_W = \hat{1}_4$ are hypercharge
operators;
$\sigma_k$ ($k = 1,\,2,\,3$) are standard ($2\times2$)
Pauli matrices; and $g^{\prime}$ and $g^{\prime\prime}$ are interaction
constants.
Connections ${\overline\Gamma}_a$ (matrices $4\times 4$)
pertain to gravitational fields only.
${A}\otimes{B}$ denotes the Kronecker product of
matrices ${A}$ and ${B}$.
{}From structure of (\ref{cderiv3}) follows that under respective matrix
transformations components $U' = [u_1,\,u_2]$ of the scalar
$U = [u_1,\,u_2,\, u_3,\,u_4]^T$
do not mix with components $U'' = [u_3,\,u_4]$ and vice versa, which
means that one
may consider two independent doublets, $[U',\,U'']$, combined in $U$.
Dirac matrices,
$\gamma_{a}$, have the following structure. In $L$-basis,
$\gamma_{a}  = $ diag$\,\left({\overline\gamma}_a\,,
{\overline\gamma}_a\,\right)
       =  \hat1_2\otimes{\overline\gamma}_a$,
where ${\overline\gamma}_a$ are Dirac matrices $4\times4$.
In $R$-basis, as well as in scalars' $U$ and $V$ bases, $\gamma_a =
{\overline\gamma}_a$.

3. To be more specific about the gravitational sector, one may introduce four
standard Dirac matrices $4\times4$, $\Delta_A$, $A = 0, 1, 2, 3$:
\begin{eqnarray}
\Delta_0 & = & \left(\begin{array}{cc}
          0 & \hat1_2 \\
          \hat1_2 & 0 \end{array}\right)\,,\,\,\,
\Delta_k  =  \left(\begin{array}{cc}
          0 & \sigma_k\\
          - \sigma_k & 0 \end{array}\right)\,,\;  k = 1, 2, 3.\nonumber
\end{eqnarray}
Then, the following decompositions take place:
\begin{eqnarray}\label{tetrad}
{\overline\gamma}_a & = & e_a^A\Delta_A\,,\\
\label{f_fields}{\overline\Gamma}_a & = &
f_a^{AB}\Delta_{[A}\Delta_{B]}\,,
\end{eqnarray}
where $e_a^A$ is a tetrad, and $f_a^{AB} = f_a^{[AB]}$
are six vector fields, which are associated with the Ricci
rotation coefficients \cite{fok} in standard theory.\footnote{{ }Namely,
$f_a^{AB} = {1\over 4}e_b^B{\nabla}_a\,e^{Ab}$, where
${\nabla}_a$ is a covariant derivative, associated with metric
$g_{ab} = e_a^Ae_{Ab}$.}

4. One may use the scalar fields $\overline U$, $U$
in order to implement symmetry breaking mechanism for generation of
masses for vector bosons and fermions. The respective potential
energy,
\begin{equation}\label{penergy}
V_p(\overline{U},U,\overline{V},V) \,=\,
\sigma_U\, \overline{U}U + \kappa_U\, (\overline{U}U)^2 +
\sigma_V\,\overline{V}V\,,
\end{equation}
where $\sigma_U$, $\sigma_V$, and $\kappa_U$ are constants.

Expanding  $\sqrt{-\chi}$ in powers of $\ell$, one obtains,
$\sqrt{-\chi}  =  \sqrt{-g}\left( 1  - {1\over2}\,
\pi_a\pi^a \right) + O(\ell^4)$. Here  $\pi^a = g^{ab}\pi_b$.
In the minimum of potential energy (\ref{penergy}) scalar fields may be taken with
the following constant values: $U' = [0,\,p]^T$, $U'' = [0,\,q]^T$, and
$\overline{U}' = [0,\,p]$, $\overline{U}'' = [0,\,q]$;
one obtains, then,
\begin{eqnarray}\label{chiex}
(\overline{U}U)^{-1}\cdot \overline{U}\sqrt{-\chi}U & = & \sqrt{-g}  - \sqrt{-g}\,{{\ell^2}\over8}
g'^2\left(W_a^1W^{1a} + W_a^2W^{2a}\right)\nonumber\\
& & - \sqrt{-g}\,{{\ell^2}\over8}\left(g'W_a^3 - g''B_a\right)
\left(g'W^{3a} - g''B^a\right) + O(\ell^2)\,.
\end{eqnarray}
Analogously, one may write an expansion
\begin{eqnarray}\label{chiexv}
\overline{V}\sqrt{-\chi}V & = & \sqrt{-g}\left( \overline{V}V
+ {{\ell^2}\over2}\,
\overline{V}\partial_a\partial^a V \right) + O(\ell^4)\,.
\end{eqnarray}

5. Similarly, expanding (\ref{phi}) into row in powers of $\ell$,
and using (\ref{myform}), one obtains for two spinorial bases the following
decompositions. For the $L$-basis,
\begin{eqnarray}\label{exp_phi_L}
\sqrt{-\phi_L} & = & \sqrt{-g} + {1\over{48}}\sqrt{-g}\,
Tr\lbrace \overline{\gamma}^a\overline{\gamma}^b
\overline{\rho}_{ab}\rbrace \nonumber \\
&   &  + {{\ell^4}\over{320}}\sqrt{-g}\left(
g^{\prime2}W_k^{ab}W_{ab}^k + g^{\prime\prime2}B^{ab}B_{ab}\right)
+ O(\ell^4)\,.
\end{eqnarray}
For the $R$-basis,
\begin{eqnarray}\label{exp_phi_R}
\sqrt{-\phi_R} & = & \sqrt{-g} + {1\over{48}}\sqrt{-g}\,
Tr\lbrace \overline{\gamma}^a\overline{\gamma}^b
\overline{\rho}_{ab}\rbrace
+ {{\ell^4}\over{80}}\sqrt{-g}\,
g^{\prime\prime2}B^{ab}B_{ab} + O(\ell^4)\,,
\end{eqnarray}
where $\overline{\rho}_{ab}$ is constructed via $\overline{\Gamma}_a$
as in (\ref{curvature}).
Only terms pertaining to known components of the
standard model are extracted.
One uses notations $W_{ab}^k = \partial_aW_b^k -  \partial_bW_a^k -
g^{\prime}\epsilon_{lm}^kW_a^lW_b^m$
      and $B_{ab} = \partial_aB_b - \partial_bB_a$ for
electroweak gauge fields.\footnote{{ }Indices k, l, m ... are raised
and lowered with $\delta_{km} = \delta^{km} = diag (1, 1, 1)$.}
As shown below, the second term in the right hand side of (\ref{exp_phi_L})
and  (\ref{exp_phi_R}) corresponds to EH term.

As it follows from (\ref{phiop}), (\ref{phi}), and (\ref{cderiv4}),
\begin{equation}\label{phi_V}
\sqrt{-\phi_{V}}\, =\, \sqrt{-g}\,,
\end{equation}
where $\phi_{V}$ denotes $\phi$ (c.f. (\ref{phi})) given in
$V$-representation.

6. Comparing (\ref{chiex}) - (\ref{phi_V}), one defines 
action for the fields,
\begin{eqnarray}\label{S_f}
S_f & = & \int d\Omega \nonumber\\
& & \times
\left[\lambda_L\sqrt{-\phi_L} + \lambda_R\sqrt{-\phi_R} -
  \Upsilon\,\sqrt{-\phi_{V}} +
\lambda_{U}\overline{U}\sqrt{-\chi}U  +
\lambda_{V}\overline{V}\sqrt{-\chi}V\right]\,,
\end{eqnarray}
where one introduces  constants,
$\lambda_L, \lambda_R, \lambda_{U}, \lambda_{V}$,
and $\Upsilon$, containing the potential (\ref{penergy}),
necessary for symmetry-breaking mechanism:
\begin{eqnarray}\label{potential}
\Upsilon & = &\lambda_L + \lambda_R  + V_p\,.
\end{eqnarray}
The action (\ref{S_f}) together with the action for spinors
((\ref{S_psi}) below) should be varied with respect to fields
$e_a^A,\,f_a^{AB},\,W_a^k,\,B_a,\,\overline{U},\, U,\,
\overline{V},\, V ,\,
\overline{L}_{el}$, $L_{el}$, $\overline{e}_R$, and $e_R$.

Comparing the expansion of (\ref{S_f}) in $\ell$ with the standard action,
one obtains the following expressions for constants:
\begin{eqnarray}
\label{ell}
\ell^2 & = & {{10}\over{3\alpha}}(1 + 2\sin^2\theta_W)\,\ell_p^2\,;\\
\label{lambda_L}
\lambda_L & = & - {6\over{\pi}}\,{{\sin^2\theta_W}\over
{1 + 2\sin^2\theta_W}}\,{{c^3}\over{k\ell^2}}\,;\\
\label{lambda_R}
\lambda_R & = & - {3\over{2\pi}}\,{{1 - 2\sin^2\theta_W}\over
{1 + 2\sin^2\theta_W}}\,{{c^3}\over{k\ell^2}}\,;\\
\label{lambda_U}
\lambda_{U} & = & - {1\over{8\pi c\ell^2}}\,;\\
\label{lambda_V}
\lambda_{V} & = & - {1\over{8\pi c\ell^2}}\,;\\
\label{mass}
p^2 + q^2 & = & 4\sin^2\theta_W\,{{m_W^2c^4}\over{e^2}}\,,
\end{eqnarray}
where $\alpha$ is the fine structure constant for an electron,
$\theta_W$ is Weinberg's mixing angle \cite{Halzen},
$\ell_p$ is Planck's length, and $m_W$ is vector bosons' rest mass.

Note, that scalar field's $V$ sole purpose is to make invariant (\ref{phi_V}) 
possible; in turn, one needs $\sqrt{-\phi_V}$ in order to adjust the `cosmological term', so
that it would have a reasonable value.
The field $V$ doesn't interact with other fields. On the other hand, it contributes stress-energy
to the Einstein equations; thus one may think of it as of the `dark matter'.

\section{Action for spinors}

The relevant tensors, invariant with respect to (\ref{symm}) and
(\ref{symm1}), are:
\begin{eqnarray}\label{contra}
{\phi}^{dh} & = &
{4\over{5!\phi}}e^{abcd}e^{efgh}\,\phi_{ae}\,\phi_{bf}\,\phi_{cg}\,;\\
\label{sigma}\Sigma_{ab} & = & \gamma_{[a}\pi_{b]} - \pi_{[a}\gamma_{b]}\,.
\end{eqnarray}
Then, the action for fermions may be constructed as follows:
\begin{eqnarray}\label{S_psi}
S_{\Psi} & = & C\int d\Omega \nonumber\\
& & \times \,\left[\sqrt{-\phi_L}\,
\overline{L}_{el}\,\phi^{ab}\,\Sigma_{ab} L_{el}
+ \sqrt{-\phi_R}\,\overline{e}_R\,\phi^{ab}\,\Sigma_{ab} e_R
      +  r\sqrt{-\phi_{V}}\,\left( \overline{L}_{el}e_RU' +
\overline{U}'\overline{e}_RL_{el}\right)\right]\,.
\end{eqnarray}
Expanding (\ref{S_psi}) in powers of $\ell$, one obtains,
\begin{eqnarray}\label{d_exp}
S_{\Psi} = {{2\ell C}\over{\hbar}}\int d\Omega\sqrt{-g}\left[
\overline{L}_{el}\,\gamma^a{p}_a L_{el}
+ \overline{e}_R\,\overline{\gamma}^a{p}_a e_R
      + {{\hbar r p}\over{2\ell}}\left(
\overline{e}_Le_R + \overline{e}_Re_L\right)\right] + \cdots
\end{eqnarray}
The momentum operator, ${p}_a = -i\hbar (\partial_a - \Gamma_{a})$, is
given in respective spinorial bases.

\section{Gravitational sector}

It will be shown, that the gravitational part of
the action (\ref{S_f}) in vacuum (in the limit $\ell\rightarrow0$)
is equivalent to the EH action in general
relativity.
First, extract the action of the gravitational field from the expansion of
(\ref{S_f}) in powers of $\ell$:
\begin{eqnarray}\label{S_g}
S_g & = &
\kappa^{-1\,}{\int}d\Omega\,\sqrt{-g}\,{Tr}\left\{
{\overline{\gamma}}^a\,{\overline{\gamma}}^b\,
{\overline{\rho}}_{ab}\right\}\,,
\end{eqnarray}
Varying (\ref{S_g}) with respect to $\overline{\gamma}_a$
and $\overline{\Gamma}_a$, one obtains the
equations,
\begin{eqnarray}\label{E_for_gamma}
\left[ \overline{\gamma}^b\,, \overline{\rho}_{ab}\right]
- {1\over 4}\,\overline{\gamma}_a\,
Tr\lbrace \overline{\gamma}^c\,\overline{\gamma}^d\,
\overline{\rho}_{cd}\rbrace & = & 0\,,\\
\label{E_for_conn}
{1\over{\sqrt{-g}}}\,\partial_b\left(
\sqrt{-g}\,\overline{\gamma}^{[a}\,\overline{\gamma}^{b]}\right)
- \left[ \overline{\Gamma}_b\,, \overline{\gamma}^{[a}\,
\overline{\gamma}^{b]} \right] & = & 0\,.
\end{eqnarray}
Equations (\ref{E_for_conn}) are equivalent to the following:
\begin{eqnarray}\label{D_gamma2}
\left( \overline{\gamma}^{[a}\,\overline{\gamma}^{b]}
\right)_{\;;\,b} & = & 0\,,
\end{eqnarray}
where the semicolon denotes covariant derivative.
For the covariant derivative of $\overline{\gamma}_a$ one obtains,
\begin{eqnarray}\label{D_gamma_def}
\overline{\gamma}_{a\, ;\, b} & = & {i\over {\ell}}\,[\pi_b, \overline{\gamma}_a] 
 \,=\, \partial_b\overline{\gamma}_{a}
- {\Gamma}_{ab}^c\overline{\gamma}_{c} -
\left[ \overline{\Gamma}_{b}\,, \overline{\gamma}_{a} \right] \,.
\end{eqnarray}
{}From (\ref{D_gamma2}) follows (as one possible solution),
\begin{eqnarray}\label{D_gamma1}
\overline{\gamma}_{a\, ;\, b} & = & 0\,.
\end{eqnarray}
Applying the antisymmetrized product of covariant derivatives
to $\overline{\gamma}_{b}$ and using (\ref{D_gamma1}), one obtains,
\begin{eqnarray}\label{riemann}
R_{\,bcd\,}^{\,a}\overline{\gamma}_{a} & = &
{1\over{\ell^2}}\lbrack\overline{\gamma}_{b}\,,
{\overline{\rho}}_{cd}\rbrack\,,
\end{eqnarray}
where $R_{\,bcd}^{\,a}$ is the Riemann tensor,
\begin{eqnarray}\label{dreim}
R_{\,bcd}^{\,a} & = & 2\left(\partial_{[c|}\Gamma_{b|d]}^a -
\Gamma_{b[c|}^e\Gamma_{e|d]}^a\right)\,.
\end{eqnarray}
For the scalar curvature and
Ricci tensor one obtains, respectively,
\begin{eqnarray}\label{scalcur}
R & = & {1\over{2\ell^2}}
{Tr}\left\{{\overline{\gamma}}^a\,{\overline{\gamma}}^b\,
{\overline{\rho}}_{ab}\right\}\,,\\
\label{ricci}R_{\,bd} & = & {1\over{4\ell^2}}
{Tr}\left\{{\overline{\gamma}}^a\lbrack\overline{\gamma}_{b}\,,
{\overline{\rho}}_{ad}\rbrack\right\}.
\end{eqnarray}
In given approximation (i.e., in the limit $\ell \rightarrow 0$),
it follows from (\ref{riemann}),
\begin{equation}\label{curvature_tensor}
\overline\rho_{cd} \,=\, - {{\ell^2}\over 4}\,R_{abcd}\,
\overline\gamma^a\overline\gamma^b \,.
\end{equation}
Here, again, $\overline\gamma^a = g^{ab}\,\overline\gamma_b$.
Multiplying (\ref{E_for_gamma}) by $\overline{\gamma}_{c}$ and
taking the trace, one obtains the vacuum Einstein equations,
\begin{eqnarray}\label{gen_rel_eq}
R_{ac} - {1\over2}g_{ac}R & = & 0\,,
\end{eqnarray}
where definitions (\ref{ricci}) and (\ref{scalcur}) are used
for the Ricci tensor and scalar curvature in
\hbox{`$\overline{\rho}$--$\overline{\gamma}$'}
representation. Note that on the classical level, when the matter
lagrangian doesn't depend on $\overline{\Gamma}_{a}$,
equations (\ref{gen_rel_eq})
are still valid, if one adds the stress-energy tensor for matter to the
right hand side of (\ref{gen_rel_eq}).

\section{Meaning of parameter $\theta$}

{}From the structure of the action for the electron (\ref{d_exp}), follow
expressions for the correction to the energy, $\delta E$, due to the value
of the potentials of electroweak fields at spatial infinity.
Namely,
\begin{eqnarray}\label{delta_e1}
\delta E_L &=& {\hbar\over 2}\left(g'W_0^3(\infty) + g''B_0(\infty)\right)
\int dV\, \overline{e}_L\overline\gamma^0 e_L\,;\\
\label{delta_e2}
\delta E_R &=& \hbar g''B_0(\infty) \int dV
\,\overline{e}_R\overline\gamma^0 e_R\,.
\end{eqnarray}
One uses notations, $W_0^3(x^a) \rightarrow W_0^3(\infty)$,
as $|x^a| \rightarrow \infty$, etc.
{}From (\ref{symm}) and  (\ref{symm1}) follows,
\begin{eqnarray}\label{symm_int}
\int \overline{L}_{el}{\gamma'}^a  d\Sigma_a L_{el}& = &
\cosh\theta\,\int \overline{L}_{el}\gamma^a  d\Sigma_a L_{el}
+ \sinh\theta\, \int \overline{L}_{el}\pi^a d\Sigma_a L_{el}\,,\\
\label{symm1_int}\int \overline{L}_{el}{\pi'}^a  d\Sigma_a L_{el}
& = &
\sinh\theta\,\int \overline{L}_{el}\gamma^a  d\Sigma_a L_{el}
+ \cosh\theta\,\int \overline{L}_{el}\pi^a  d\Sigma_a L_{el}\,.
\end{eqnarray}
One uses notations, ${\gamma'}^a = \chi^{ab}{\gamma'}_b$, $\gamma^a = \chi^{ab}\gamma_b$,
${\pi'}^a = \chi^{ab}{\pi'}_b$, and ${\pi}^a = \chi^{ab}{\pi}_b$
One integrates over an arbitrary hypersurface with $d\Sigma_a = $
$\sqrt{-\chi}\,e_{abcd}\,dx^b\,d{x'}^c\,d{x''}^d$.
One may denote energy, $E = \int \overline{L}_{el}\ p^0 dV L_{el} $,
and electrical charge, $Q = \int \overline{e}_L\overline\gamma^0 dV e_L$,
where $dV \equiv d\Sigma_0$ is the element of spatial volume, and
$p^a = {\hbar\over\ell}\,\pi^a$ is the momentum operator.
Selecting spatial volume as a hypersurface of integration,
one obtains
\begin{eqnarray}\label{E}
E' &=& \cosh\theta\, E + {\hbar\over\ell}\,\sinh\theta\, Q\,;\\
\label{Q}
Q' &=& \cosh\theta\, Q + {\ell\over\hbar}\, \sinh\theta\,E\,.
\end{eqnarray}
Equations (\ref{E}) and (\ref{Q}) represent transformations
of energy and electrical charge, respectively. Suppose, that
$E = 0$. Then, $\tanh\theta = {\ell\over\hbar}\,{{E'}\over{Q'}}$,
which means
that energy $E'$ is due as a whole to electrical charge $Q'$,
placed in constant electroweak potential, as in (\ref{delta_e1}).
Thus,
\begin{equation}\label{e_L}
\tanh \theta \,=\, {\ell\over 2}\left(g'W_0^3(\infty)
+ g''B_0(\infty)\right)\,.
\end{equation}
Analogous considerations for $e_R$ instead of $L_{el}$ lead to the equation
\begin{equation}\label{e_R}
\tanh \theta \,=\, \ell g'' B_0(\infty)\,,
\end{equation}
{}from which it follows that one should put $g'W_0^3(\infty)
= g''B_0(\infty)$.\footnote{{ }One should use
$\sin\theta_W\,g' = \cos\theta_W\,g'' = {e\over{\hbar c}}$.}
Thus, one should associate the $\theta$-transformation with that
of electroweak potentials at spatial infinity.
{}From (\ref{e_L}) it follows that $\theta \propto \ell$,
and neglecting terms of order $\ell^2$ in  (\ref{Q}) , one obtains $Q' \approx Q$.



\end{document}